\documentclass[12pt]{article}
\usepackage[margin=1.75in]{geometry}
\usepackage{biblatex}
\usepackage{physics}
\usepackage{amsmath}
\begin{document}
\title{\textbf{Perturbation of Quantum Harmonic Oscillator and its effect on Quantum Electromagnetic Field Theory}}
\author{Sankarshan Sahu}
\maketitle
\begin{abstract}
\textit{Study of perturbation theory has been quite popular in quantum mechanics. One of the most important systems that is very crucial to the framework of quantum field theory is the system of harmonic oscillator.  Here a special case of perturbation in quantum harmonic oscillator is studied. Here we assume the perturbed potential to be a Harmonic Oscillator that has been shifted in the position space.We construct the new creation and annihilation operators for the new Hamiltonian to find out its energy eigenstates. What is interesting about the solution of this system is that we find out that the ground state of this system is a coherent state of the ground state of the original harmonic oscillator. We know that the Hamiltonian of electromagnetic field  is similar to that of the harmonic oscillator where the electric field and magnetic field are expressed in terms of time evolution of annihilation and creation operators. According to the quantum electromagnetic field theory, photon states are expressed as the coherent states of the ground states of the unperturbed harmonic oscillator.In the previous research works,it has been established how coherent states provide a better picture of Electric field compared to the energy eigenstates. This paper mainly deals with why the coherent states provide a more complete picture of electric and magnetic field theory compared to just energy eigenstates. Here we find how our perturbation act as a coherent state generator and adds a new term in the Hamiltonian. The Perturbation is thought to arise from the only available interaction i.e. observation. The perturbation also explains how observation leads to coherence of photon states. We will find out how the new perturbation affects the time evolution of electromagnetic field. This also explains how the classical electro-magnetic field theory takes into account the factor of observation, due to which the classical laws are in perfect agreement with the quantum laws.}
\end{abstract}
\section{Introduction}
Coherent states already appeared in physics in the 1920s, when
Schrödinger noticed the existence of superpositions of quantum
states that exhibit various features like dynamical behaviours quite
similar to those of the classical counterparts. In this sense, coherent
states constitute an interesting tool that allows to reproduce,
under some constraints, properties of the classical behaviour of
the system. Although initially this important observation passed
unnoticed, the notion of coherent states was rediscovered in the
frame of quantum optics in 1963. From that time onwards, coherent
states have been recognized as an essential concept in many
different domains of physics, and various types of generalizations
have been proposed.
\\Quantum field theory is a theory in which all the existing fields are expressed as operators and the existing particles are visualized as ripples in these fields, as if the localization of the field gives rise to particles. In the quantum electromagnetic field theory, the electric and magnetic fields are expressed as operators, and the particles which are assumed to be the localization of this field are photons.
\\The quantum harmonic oscillator is one of the most important system in the framework of quantum mechanics. The Schrodinger equation for Harmonic oscillator can be solved by two distinct ways. However in our discussion ahead we will be talking about the analytical solution of the Schrodinger’s equation through ladder operators which provides a deeper insight to the time evolution of the operators which will be useful farther in our analysis of fields. This method uses the “creation and annihilation” operators to produce quantization of the energy eigenstates. The mathematical backdrop for understanding the work ahead shall be discussed later.
\\Coherent states are those states which moves in space without changing its shape. The statement above shall seem vague but with required mathematical backdrop it is fairly easy to understand and visualize the coherent states which will form the basis of our discussion ahead.
\\The previous works done on this topic suggests that  the Hamiltonian associated with the electric and magnetic fields is the harmonic oscillator. The energy eigenstates of that oscillator are called the photon states. The classical electric field is often expressed as the expectation value of the electric field operator in the coherent state of these photon states. My work explains how these coherent states emerge out of a new Hamiltonian produced as a result of the interaction between the electric and magnetic fields. We will find that the new Hamiltonian is nothing but a perturbation to our old Hamiltonian whose energy eigentates are the coherent sates of photon states. However one thing one must keep in his/her mind is that the Quantum field theory assumes the fields as the more fundamental frameworks which give rise to particles, and not the other way around. So our Classical intuition may not always hold true.
\\In this paper mainly Dirac's Formalism has been used.
\section{Quantum Harmonic Oscillator}
The Hamiltonian of the Quantum Harmonic Oscillator can be written as:-
\begin{equation}\tag{2}
\hat{H}=\frac{\hat{p}^2}{2m}+\frac{1}{2}m\omega^2{\hat{x}}^2
\end{equation}
where, $\hat{p}$ represents the momentum operator and $\hat{x}$ represents the position operator. The mass the of the prticle is represented by 'm'.
The operators $\hat{p}$ and $\hat{x}$ are canonical operators.The observables $\hat{x}$ and $\hat{p}$ don't commute.The commutation relation between them is shown below:-
$$\comm{\hat{x}}{\hat{p}}=i\hbar$$
The solution of the time-independent Schr\"{o}dinger's equation for the above Hamiltonian can be done in two ways. One way is finding approximate mathematical solution to this differential equation. However, the solution we will be considering here is the ananalytical solution. Here we use the creation and annahilation operators to raise and lower the eigenstates.These operators are also sometimes reffered to as ladder operators.
\\The annahilation and creation operators are described as:-
\begin{equation}\tag{2.1a}
\hat{a}=\sqrt{\frac{m\omega}{2\hbar}}(\hat{x}+\frac{i\hat{p}}{m\omega})
\end{equation}
\begin{equation}\tag{2.1b}
\hat{a}^{\dagger}=\sqrt{\frac{m\omega}{2\hbar}}(\hat{x}-\frac{i\hat{p}}{m\omega})
\end{equation}
where $\hat{a}$ represents the annahilation operator and $\hat{a}^\dagger$ represents the creation operator.
The operators $\hat{a}$ and $\hat{a}^{\dagger}$ don't commute.
The commutation relation between them is given by:-
\begin{equation}\tag{2.1c}
\comm{\hat{a}}{\hat{a}^{\dagger}}=1
\end{equation}
Let $\ket{n}$ be the n-th energy eigenstate of the Hamiltonian.
Then we have:-
\\$$\hat{a}\ket{n}=\sqrt{n}\ket{n-1}$$ and
$$\hat{a}^{\dagger}\ket{n}=\sqrt{n+1}\ket{n+1}$$
The Hamiltonian of the system can be re-written just using $\hat{a}$ and $\hat{a}^{\dagger}$ operators. i.e.
$$\hat{H} = \hbar\omega(\hat{a}^{\dagger}\hat{a}+\frac{1}{2})$$
The energy eigenvalues of the Hamiltonian are given by:-
$$ E_n=\hbar\omega(n+\frac{1}{2})$$
where n represents the n-th energy eigenstate of the Hamiltonian.
From this, we understand that no matter what,the expectation value of $\hat{a}^{\dagger}\hat{a}$ will always be a pure number.
i.e. $\bra{n}\hat{a}^{\dagger}\hat{a}\ket{n} = n$ where n is a pure number(in this case, n is the n-th energy eigenstate.) Thus the operator $\hat{a}^{\dagger}\hat{a}$ is often called as the number operator. The number operator is represented by $\hat{N}$.
So we have:-
$$\hat{N}=\hat{a}^{\dagger}\hat{a}$$
So far, we have devoloped the required mathematical model for understanding the Harmonic Oscillator. Now we shall look at our case of the Perturbed Hamiltonian and look at the various changes we need to consider when perturbation is done to the system.
\section{Time evolution of operators}
From the Heisenberg picture of Quantum Mechanics we know that:-
\begin{equation}\tag{3.1}
i\hbar\dfrac{\hat{dA_H}}{dt}=\comm{\hat{A}_H}{\hat{H}_H}+i{\hbar}\left(\frac{\partial \hat{A_s}}{\partial t}\right)_H
\end{equation}
Here the subscripts H and S represents The Heisenberg and the Schr\"{o}dinger operators respectively. Thus $\hat{A}_H$ represents the heisenberg operator whereas, the operator $\hat{A}_s$ represents the schr\"{o}dinger operator.
The heisenberg operators of the corrosponding Schr\"{o}dinger operators are nothing but the time evolution of the Schr\"{o}dinger operators.
\\ The Heisenberg operators can be expressed in terms of The Schr\"{o}dinger operators in the following way:-
$$\hat{A}_H(t)=U^{\dagger}(t,0)(\hat{A}_s)U(t,0)$$
\\ where U(t,0) represents an unitary time operator and $U^{\dagger}$(t,0) represents its complex conjugate.
Since, $\hat{A}_s$ represents th Schr\"{o}dinger operator it's Partial derivative is equal to 0. Equation 3.1 can be written as:-
$$i\hbar\frac{d\hat{A}_H}{dt}=\comm{\hat{A}_H}{\hat{H}_H}$$
This equations can be used to find out the time evolution of momentum and position operators in case of a Harmonic oscillator.
Using this equation, we find that :-
\begin{equation}\tag{3.2a}
\hat{x}_H(t)=\hat{x}cos({\omega}t)+\frac{\hat{p}}{m{\omega}}sin({\omega}t)
\end{equation}
\begin{equation}\tag{3.2b}
\hat{p}_H(t)=\hat{p}cos({\omega}t)-m{\omega}\hat{x}sin({\omega}t)
\end{equation}
\\Note:- Here the Schr\"{o}dinger operators  are not followed by the subscript s.
So the heisenberg Hamiltonian can be written as:-
$$\hat{H}_H(t)=\frac{\hat{p}_H^2(t)}{2m}+\frac{m\omega^2{\hat{x}_H}^2(t)}{2}$$
Now putting values of $\hat{p}_H(t)$ and $\hat{x}_H(t)$ from equations (3.2a) and (3.2b)in the above expression we find that:-
\begin{equation}\tag{3.3c}
\hat{H}_H(t)=\hat{H}_s
\end{equation}
Using this we can find the time evolution of the annahilation and creation operator.
Using equation(3.3) We find that:-
\begin{equation}\tag{3.4a}
\hat{a}_H(t)=\hat{a}e^{-i{\omega}t}
\end{equation}
\begin{equation}\tag{3.4b}
\hat{a}^{\dagger}_H(t)=\hat{a}^{\dagger}e^{i{\omega}t}
\end{equation}
\section{Coherent states}
Coherent states are those states(of Harmonic Oscillator) which moves in space without changing its shape at any point in time. The generalized expression of Coherent states can be expressed as :-
\begin{equation}\tag{4.1}
\ket{\alpha}=D(\alpha)\ket{0}
\end{equation}
,where $D(\alpha)$ is known as the displacement operator.
\begin{equation}\tag{4.2}
D(\alpha)\equiv exp(\alpha\hat{a}^{\dagger}-\alpha^{*}\hat{a})
\end{equation}
Here, $\hat{a}^{\dagger}$ and $\hat{a}$ are the annahilation and creation operators respectively. 
\\\textbf{Action of annahilation operator on Coherent states}
\\$$\hat{a}\ket{\alpha}=\hat{a}e^{\alpha\hat{a}^{\dagger}-\alpha^{*}\hat{a}}\ket{0}=\comm{\hat{a}}{e^{\alpha\hat{a}^{\dagger}-\alpha^{*}\hat{a}}}\ket{0}=\comm{\hat{a}}{{\alpha\hat{a}^{\dagger}-\alpha^{*}\hat{a}}}e^{\alpha\hat{a}^{\dagger}-\alpha^{*}\hat{a}}\ket{0}=\alpha\ket{\alpha}$$
\\ Thus we conclude that:-\begin{equation}\tag{4.3}
\hat{a}\ket{\alpha}=\alpha\ket{\alpha}
\end{equation} 
\section{Quantum Electromagnetic Field Theory}
For a classical electromagnetic field the energy E is obtained by adding the contributions
of the electric and magnetic field:
$$E=\int d^{3}x\:\frac{1}{2}\epsilon_{0}\;[\textbf{E}^2(r,t)+c^2\textbf{B}^2(r,t)]$$
,where \textbf{E}(r,t) and \textbf{B}(r,t) represents the Electric and Magnetic field respectively.\\We consider a rectangular cavity of volume V with a single mode of the electromagnetic field,
namely, a single frequency $\omega$ and corresponding wavenumber $k={\omega}/c$.The electromagnetic fields form a standing wave in which electric and magnetic fields are out of phase. They can take the form:-
\begin{equation}\tag{5a}
\textbf{E}_x(z,t)=\sqrt{\frac{2}{V\epsilon_0}}{\omega}{q(t)}\sin{\textit{kz}}
\end{equation}
\begin{equation}\tag{5b}
c\textbf{B}_y(z,t)=\sqrt{\frac{2}{V\epsilon_0}}{p}(t)\cos{\textit{kz}}
\end{equation}
The classical time-dependent functions q(t) and p(t) are to become in the quantum theory
Heisenberg operators $\hat{q}(t)$ and $\hat{p}(t)$ with commutation relations $\comm{\hat{p}}{\hat{q}}=i\hbar$
A calculation of the energy E in the above equation with the fields above gives:-
$$E=\frac{1}{2}(p^2(t)+{\omega}^2q^2(t))$$
There is some funny business here with units. The variables q(t) and p(t) do not have
their familiar units, as you can see from the expression for the energy. Indeed one is
missing a quantity with units of mass that divides the $p^2$ contribution and multiplies the
$q^2$ contribution. In fact, p has units of $\sqrt{E}$ and q has units of $T\sqrt{E}$. Still, the product of
q and p has the units of $\hbar$, which is useful. Since photons are massless particles there is no
quantity with units of mass that we can use. Note that the dynamical variable q(t) is not a
position, it is essentially the electric field. The dynamical variable p(t) is not a momentum,
it is essentially the magnetic field.
\\The quantum theory of this EM field uses the structure implied by the above classical
results. From the energy above we are let to postulate a Hamiltonian:-
\begin{equation}\tag{5.1}
\hat{H}=\frac{1}{2}(\hat{p}^2+{\omega}^2\hat{q}^2)
\end{equation}
\\with Schr\"{o}dinger operators $\hat{q}$ and $\hat{p}$ (and associated Heisenberg operators $\hat{q}(t)$ and $\hat{p}(t)$
that satisfy $\comm{\hat{q}}{\hat{p}} = i\hbar$ As soon as we declare that the classical variables q(t) and p(t)
are to become operators, we have the implication that the electric and magnetic fields in the equation will become field operators, that is to say, space and time-dependent operators (more
below!). This oscillator is our familiar SHO, but with m set equal to one, which is allowed
given the unusual units of $\hat{q}$ and $\hat{p}$. With the familiar (2.1(a,b,c)) and \\m = 1 we have:-
\begin{equation}\tag{5.2a}
\hat{a}=\frac{1}{\sqrt{2\hbar\omega}}({\omega}\hat{q}+i\hat{p})
\end{equation}
\begin{equation}\tag{5.2b}
\hat{a}^{\dagger}=\frac{1}{\sqrt{2\hbar\omega}}({\omega}\hat{q}-i\hat{p})
\end{equation}
\begin{equation}\tag{5.2c}
\comm{\hat{a}}{\hat{a}^{\dagger}}=1
\end{equation}
It follows that:-
\begin{equation}\tag{5.3}
\hbar\omega\hat{a}^{\dagger}\hat{a}=\\\frac{1}{2}(\omega\hat{q}-i\hat{p})(\omega\hat{q}+i\hat{p})=\\\frac{1}{2}(\hat{p}^2+{\omega}^2\hat{q}^2+i\omega\comm{\hat{q}}{\hat{p}})=\\\frac{1}{2}(\hat{p}^2+{\omega}^2\hat{q}^2-\hbar\omega)
\end{equation}
and comparing with (5.1) this gives the Hamiltonian:-
\begin{equation}\tag{5.4}
\hat{H}=\hbar\omega\left(\hat{a}^{\dagger}\hat{a}+\frac{1}{2}\right)=\hbar\omega\left(\hat{N}+\frac{1}{2}\right)
\end{equation}
This was the expected answer: this formula does not depend on m and our setting m = 1 had no import. At this point we got photons: We interpret the state $\ket{n}$ of the above harmonic oscillator as the state with n photons. This state has energy $\hbar\omega(n + \frac{1}{2})$which is, up to the zero-point energy $\hbar\omega/2$, the energy of n photons each of energy $\hbar\omega$. A
photon is the basic quantum of the electromagnetic field.For more intuition we consider the electric field operator. For this we first note that:-
\begin{equation}\tag{5.5}
\hat{q}=\sqrt{\frac{\hbar}{2\omega}}(\hat{a}+\hat{a}^{\dagger})
\end{equation}
And the corresponding Heisenberg operator is, using equations (3.4a) and (3.4b)
\begin{equation}\tag{5.6}
\hat{q}(t)=\sqrt{\frac{\hbar}{2\omega}}(\hat{a}e^{-i{\omega}t}+\hat{a}^{\dagger}e^{i{\omega}t})
\end{equation}
In quantum field theory, the electric field is a Hermitian
operator. Its form is obtained by substituting (5.6) into (5a),we have:
\begin{equation}\tag{5.7}
\hat{E_x}(z,t)=\varepsilon_0(\hat{a}e^{-i{\omega}t}+\hat{a}^{\dagger}e^{i{\omega}t})\sin {\textit{kz}} 
\end{equation}
[  where,
$\varepsilon_0=\sqrt{\frac{\hbar\omega}{{\epsilon_0}V}}$ ]
\\This is a field operator: an operator that depends on time and on position, in this case z.
The constant $\varepsilon_0$ is sometimes called the electric field of a photon.\\A classical electric field can be identified as the expectation value of the electric field
operator in the given photon state. We immediately see that in the n photon state $\ket{n}$ the expectation value of $\hat{E_x}$ vanishes! Indeed,
\begin{equation} \tag{5.8}
\bra{n}\hat{E_x}(z,t)\ket{n}= \varepsilon_0(\bra{n}\hat{a}\ket{n}e^{-i{\omega}t}+\bra{n}\hat{a}^{\dagger}e^{i{\omega}t}\ket{n})\sin{\textit{kz}}=0
\end{equation}
since the matrix elements on the right hand side are zero. Energy eigenstates of the photon field do not correspond to classical electromagnetic fields. Consider now the expectation
value of the field in a coherent state $\ket{\alpha}$. This time we get:-
\begin{equation}\tag{5.9a}
\bra{\alpha}\hat{E_x}(z,t)\ket{\alpha}= \varepsilon_0(\bra{\alpha}\hat{a}\ket{\alpha}e^{-i{\omega}t}+\bra{\alpha}\hat{a}^{\dagger}e^{i{\omega}t}\ket{\alpha})\sin{\textit{kz}}
\end{equation}
Since $ \hat{a}\ket{\alpha}=\alpha\ket{\alpha}$,
\begin{equation}\tag{5.9b}
\bra{\alpha}\hat{E_x}(z,t)\ket{\alpha}=\varepsilon_0(\alpha e^{-i{\omega}t}+\alpha^{*}e^{i{\omega}t})\sin{\textit{kz}}
\end{equation}
This now looks like a familiar standing wave.If we set $\alpha=|\alpha|e^{i\theta}$,we have:-
\begin{equation}\tag{5.10}
\bra{\alpha}\hat{E_x}(z,t)\ket{\alpha}=2|\alpha|\varepsilon_0\textbf{Re}({\alpha}e^{-i{\omega}t})\sin\textit{kz}=2\varepsilon_0|\alpha|\cos({\omega}t-\theta)\sin{\textit{kz}}
\end{equation}
Coherent photon states give rise to classical electric fields.It is the coherent states in which the classical electrical and Magnetic fields are recovered.Thus the study of these coherent states form the very basis of the field theory. Now we are going to see the Hamiltonian associated with these coherent states and how it affects  field theory.
\section{Perturbation of the Hamiltonian of the \\Quantum Harmonic Oscillator}
Here, We are going to see a linearized perturbation in the position 'x' of the Hamiltonian. So let's consider a Hamiltonian :-
$$H_{\alpha}=\frac{p^2}{2m}+\frac{1}{2}m{\omega}^2(x-\alpha)^2$$
So the Hamiltonian operator can be written as:-
\begin{equation}\tag{6.1}
\hat{H}_{\alpha}=\frac{\hat{p}^2}{2m}+\frac{1}{2}m{\omega}^2(\hat{x}-\alpha)^2
\end{equation}
(Here, $\alpha$ belongs to a complex number.)
One must appreciate the beauty in Quantum Mechanics, as it allows us to use the complex potentials. This can't be incorporated in classical mechanics, since complex potentials have simply no meaning in classical mechanics.
The solution to our perturbed Hamiltonian can be obtained in a similar way. Here also, we will use the creation and annahilation operators to find out the solution to the equation.
Let,
$$\hat{x}_{\alpha}=\hat{x}-\alpha$$
We can also express the momentum operator $\hat{p}$ in terms of $x_{\alpha}$
Since, $$d(x-\alpha)=dx$$; The momentum operator $\hat{p}$ remains the same, i.e.:-
$$\hat{p}=\frac{\hbar}{i}\frac{\partial}{\partial {x_{\alpha}}}=\frac{\hbar}{i}\frac{\partial}{\partial x}$$
So eqn(6.1) can be written as :-
\begin{equation}\tag{6.2}
\hat{H}_{\alpha}=\frac{\hat{p}^2}{2m}+\frac{1}{2}m{\omega}^2\hat{x}_{\alpha}^2
\end{equation}
Thus the new creation operator $\hat{a}_{\alpha}$ and new annahilation operator $\hat{a}^{\dagger}_{\alpha}$ can be expressed as:-
\begin{equation}\tag{6.3a}
\hat{a}_{\alpha}=\sqrt{\frac{m\omega}{2\hbar}}(\hat{x}_{\alpha}+\frac{i\hat{p}}{m\omega})
\end{equation}
\begin{equation}\tag{6.3b}
\hat{a}^{\dagger}_{\alpha}=\sqrt{\frac{m\omega}{2\hbar}}(\hat{x}_{\alpha}-\frac{i\hat{p}}{m\omega})
\end{equation}
Also, we will have similar commutation relations, i.e.
\begin{equation}\tag{6.3c}
\comm{\hat{a}_{\alpha}}{\hat{a}^{\dagger}_{\alpha}}=1
\end{equation}
The new annahilation and creation operators can also be expressed in terms of the old creation and annahilation operators respectively.
By replacing $\hat{x}_\alpha$ in equations (6.3a) and (6.3b) with $(\hat{x}-\alpha)$ we have:-
\begin{equation}\tag{6.4a}
\hat{a}_{\alpha}=\hat{a}-\sqrt{\frac{m{\omega}}{2\hbar}}\alpha
\end{equation}
\begin{equation}\tag{6.4b}
\hat{a}^{\dagger}_{\alpha}=\hat{a}^{\dagger}-\sqrt{\frac{m{\omega}}{2\hbar}}\alpha
\end{equation} 
Let the ground state of the Perturbed oscillator be $\ket{0_{\alpha}}$.
Thus:-
\begin{align}\tag{6.5}
\hat{a}_{\alpha}\ket{0_{\alpha}}=0\\
\Rightarrow {\left(\hat{a}-\sqrt{\frac{m\omega}{2\hbar}}\alpha\right)\ket{0_{\alpha}}}=0\tag{6.6}\\
\Rightarrow{\hat{a}\ket{0_\alpha}=\sqrt{\frac{m\omega}{2\hbar}}\alpha\ket{0_\alpha}}\tag{6.7}
\end{align}
So the state $\ket{0_{\alpha}}$ is an eigenstate of the (old)annahilation operator.Also if we set,
$\alpha\longrightarrow\alpha\sqrt{\frac{2\hbar}{m\omega}}$, the equation (6.7) becomes:-
\begin{equation}\tag{6.8}
\hat{a}\ket{0_{\alpha}}=\alpha\ket{0_\alpha}
\end{equation}
This is exactly the same equality we obtained for coherent state $\ket{\alpha}$ in equation (4.3).
So we can say that:-
\begin{equation}\tag{6.9}
\ket{0_{\alpha}}=c\ket{\alpha},
\end{equation}
Where c is any complex constant. On setting $\alpha\longrightarrow\alpha\sqrt{\frac{2\hbar}{m\omega}}$, Perturbed Hamiltonian becomes:-
\begin{equation}\tag{6.10}
\hat{H}_{\alpha}=\frac{\hat{p}^2}{2m}+\frac{1}{2}m{\omega}^2\left(\hat{x}-\sqrt{\frac{2\hbar}{m\omega}}\alpha\right)^2
\end{equation}
Since, $\ket{0_{\alpha}}$ is an eigenstate,of this Hamiltonian and $\ket{0_{\alpha}}=c\ket{\alpha}$, So we can say that, $\ket{\alpha}$ is a ground state of the Hamiltonian.
Thus, We have found a Hamiltonian whose ground state is a coherent state of the eigenstates of the Harmonic oscillator. We will use this Perturbed Oscillator in the future to construct a new approach to electromagnetic theory.
\section{Time evolution of the new annahilation and creation operators}
The time evolution of the new annahilation and creation operators can be derived from eqn(3.1). However, such rigorous calculations won't be needed in this case, since we have already calculated the time evolution of the original operators. The new Hamiltonian and the new operators are nothing but the old operators with a shift in x-space. Since this shift is by a complex constant and is independent of 'x' or any variable, the time evolution of these operators will be in a similar way as in eqns (3.4a) and (3.4b).
Thus the new time evolution of these operators will be:-
\begin{equation}\tag{7.1a}
{\hat{a}_{\alpha}}(t)=\hat{a}_{\alpha}e^{-i{\omega}t}
\end{equation}
\begin{equation}\tag{7.1b}
{\hat{a}^{\dagger}_{\alpha}}(t)=\hat{a}^{\dagger}_{\alpha}e^{i{\omega}t}
\end{equation}
\section{Effect of the Perturbation in Quantum Electromagnetic Field Theory}
In section 5 we saw that the Hamiltonian for the electro-Magnetic fields is similar to Quantum Harmonic Oscillator with mass set to 1 (m=1 in eqn(2).)
We found out that coherent states give rise to the classical electric fields. We have now found out a Hamiltonian whose ground state is the coherent state of the Quantum Harmonic oscillator. Taking this Perturbed oscillator as our new Hamiltonian for electro-magnetic fields, we are now going to calculate the expectation value of the new electric field.
Setting m=1 in equation(6.10), we get:-
\begin{equation}\tag{8.1}
\hat{H}_{\alpha}=\frac{\hat{p}^2}{2}+\frac{1}{2}{\omega}^2\left(\hat{x}-\sqrt{\frac{2\hbar}{\omega}}\alpha\right)^2
\end{equation}
The position operator $\hat{x}$ is similar to the electric field operator $\hat{q}$ and the momentum operator $\hat{p}$ is similar to the magnetic field operator $\hat{p}$, according to the quantum electromagnetic field theory.
Thus the perturbed oscillator which can be used as the Hamiltonian for the electromagnetic field is:-
\begin{equation}\tag{8.2}
\hat{H}_{\alpha}=\frac{\hat{p}^2}{2}+\frac{1}{2}{\omega}^2\left(\hat{q}-\sqrt{\frac{2\hbar}{\omega}}\alpha\right)^2
\end{equation}
Once again, we set $$\hat{q}_{\alpha}=\hat{q}-\sqrt{\frac{2\hbar}{\omega}}\alpha$$ and $$\hat{p}=\hat{p}_{\alpha}$$.
Thus the new Hamiltonian for the electromagnetic field, can be written as:-
\begin{equation}\tag{8.3}
\hat{H}_{\alpha}=\frac{\hat{p}^2_{\alpha}}{2}+\frac{1}{2}{\omega}^2\hat{q}_{\alpha}^2 
\end{equation}
The electric field operator $\hat{E_x}(z,t)$ will be defined as:-
\begin{equation}\tag{8.4}
\hat{E_x}(z,t)=\varepsilon_0\sqrt{\frac{2\omega}{\hbar}}\hat{q}(t)
\end{equation}

Now, we shall calculate the expectation value of the electric field $\hat{E}$ in the coherent states. For his, we need to calculate the value of time evolution of $\hat{q}$ as per given in section 3. According to section 3 we have:-
\begin{equation}\tag{8.5}
i\hbar\frac{d\hat{q}_{H}}{dt}=\comm{\hat{q}_H}{\hat{H}_{H}}
\end{equation}.
This calculation is messy. So we apply another technique. Since we know that $\hat{q}(t)$ can be broken down as sum of the annahilation and creation operators. So we calculate the time evolution of annahilation and creation operators. For doing so, first we shall calculate the time evolution of $\hat{a}_{\alpha}$  and $\hat{a}^{\dagger}_{\alpha}$. Since We have transformed the Hamiltonian and every other operator in terms of new '$\alpha$' co-ordinates,The time evolution of $\hat{q}_{\alpha}$ can be written as in equation 5.6.
So $\hat{q}_{\alpha}(t)$ can be written as :-
\begin{equation}\tag{8.6}
\hat{q}_{\alpha}(t)=\sqrt{\frac{\hbar}{2\omega}}(\hat{a}_{\alpha}e^{-i{\omega}t}+\hat{a}^{\dagger}_{\alpha}e^{i{\omega}t})
\end{equation}
We have derived earlier that :-
\begin{equation}\tag{8.7a}
\hat{a}_{\alpha}=\hat{a}-\alpha
\end{equation}
\begin{equation}\tag{8.7b}
\hat{a}^{\dagger}_{\alpha}=\hat{a}^{\dagger}-\alpha
\end{equation}
Plugging this in equation 8.6 we have:-
\begin{align}\tag{8.7c}
\hat{q}_{\alpha}(t)=\sqrt{\frac{\hbar}{2\omega}}((\hat{a}-\alpha)e^{-i{\omega}t}+(\hat{a}^{\dagger}-\alpha)e^{i{\omega}t})\\\tag{8.7d}
\Rightarrow \hat{q}_{\alpha}(t)=\sqrt{\frac{\hbar}{2\omega}}(\hat{a}e^{-i{\omega}t}+\hat{a}^{\dagger}e^{i{\omega}t}-\alpha(e^{-i{\omega}t}+e^{i{\omega}t}))
\end{align}
We know that:-
\begin{equation}\tag{8.8}
\hat{q}_{\alpha}(t)=U^{\dagger}(t,0)(\hat{q}_{\alpha})U(t,0)
\end{equation}
Where $U(t,0)$  represents the unitary time operator. Using this, we have:-
\begin{align*}\tag{8.9a}
& \hat{q}_{\alpha}(t)=U^{\dagger}(t,0)(\hat{q}_{\alpha})U(t,0)\\\tag{8.9b}
& \Rightarrow \hat{q}_{\alpha}(t)=U^{\dagger}(t,0)\left(\hat{q}-\sqrt{\frac{2\hbar}{\omega}}\alpha\right)U(t,0)\\\tag{8.9c}
& \Rightarrow \hat{q}_{\alpha}(t)=U^{\dagger}(t,0)(\hat{q})U(t,0)-U^{\dagger}(t,0)\left(\sqrt{\frac{2\hbar}{\omega}}{\alpha}\right)U(t,0)\\\tag{8.9d}
& \Rightarrow \hat{q}_{\alpha}(t)=\hat{q}(t)-\left(\sqrt{\frac{2\hbar}{\omega}}{\alpha}\right)U^{\dagger}(t,0)U(t,0)\\\tag{8.9e}
& \Rightarrow \hat{q}_{\alpha}(t)=\hat{q}(t)-\sqrt{\frac{2\hbar}{\omega}}\alpha
\end{align*}
(Since, $\alpha$ is a complex constant)
Putting this in equation 8.7d we have:-
\begin{align}\tag{8.10a}
\hat{q}_{\alpha}(t)=\sqrt{\frac{\hbar}{2\omega}}(\hat{a}e^{-i{\omega}t}+\hat{a}^{\dagger}e^{i{\omega}t}-\alpha(e^{-i{\omega}t}+e^{i{\omega}t}))\\\tag{8.10b}
\Rightarrow \hat{q}(t)=\sqrt{\frac{\hbar}{2\omega}}(\hat{a}e^{-i{\omega}t}+\hat{a}^{\dagger}e^{i{\omega}t}-\alpha(e^{-i{\omega}t}+e^{i{\omega}t})+2\alpha)
\end{align}
Thus we get the time evolution of the $\hat{q}$ operator. Note: It is quite different from the one we obtained in equation 5.6. This is due to the change in Hamiltonian. Since the Hamiltonian changes, its commutation relation with $\hat{q}$ also changes. Thus we get a new time evolution for $\hat{q}$. No we will calculate the expectation value of the electricfield operator in the coherent states(which is the ground state of this Hamiltonian).
So the Electric field operator can be written as:-
\begin{equation}\tag{8.11}
\hat{E_x}(z,t)=\varepsilon_0(\hat{a}e^{-i{\omega}t}+\hat{a}^{\dagger}e^{i{\omega}t}-\alpha(e^{-i{\omega}t}+e^{i{\omega}t})+2\alpha)\sin kz
\end{equation}
Its expectation value in the coherent satates is:-
\begin{align*}\tag{8.12a}
& \bra{\alpha}\hat{E_x}(z,t)\ket{\alpha}=\varepsilon_0((\alpha^{*}-\alpha)e^{i{\omega}t}+2\alpha)\sin kz\\\tag{8.12b}
& \Rightarrow \bra{\alpha}\hat{E_x}(z,t)\ket{\alpha}=\varepsilon_0(2{\alpha}-2i|{\alpha}|\sin{\theta}e^{i{\omega}t})\sin{kz}\\\tag{8.12c}
& \Rightarrow \bra{\alpha}\hat{E_x}(z,t)\ket{\alpha}=\varepsilon_0(2|{\alpha}|\cos{\theta}+2i|{\alpha}|\sin{\theta}-2i|{\alpha}|\sin{\theta}e^{i{\omega}t})\sin{kz}
\end{align*} 
(In the above calculations the value of complex constant $\alpha$ is taken as $\alpha=|{\alpha}|e^{i\theta}$) \\
Thus,We have reached a dead end in our calculations. In this case we found out a Hamiltonian whose ground state is the coherent state for the Harmonic Oscillator. Then we found out the time evolution of Electric field, for the new Hamiltonian and calculated the expectation value of the Electric field in the ground state of the new operator (which is essentially the coherent states). This expectation value however, doesn't make any sense. The expectation value should actually represent the classical electric field as in equation (5.10). But in this case, the expectation value gives out a complex expression which doesnot resemble the classical Electric-field wave equation at all.Thus, we have analysed two distinct cases. One in which, the eigenstates make the classical electric field zero and in the other one, the eigenstates give a complex expression for the electric field. The only case where we get the correct expression for the electric field is when we use the coherent states instead of eigenstates for calculation of electric field. However when we use a Hamiltonian to create these coherent states and use it to calculate the expectation value of electric field in these states, we get a complex expression which appears meaninless. This is due to the fact that, on changing the Hamiltonian the time evolution of electric field also changes. Thus we are unable to explain the emergence of coherent states using just the algebra we know.Thus, we need to find a different apporoach to explain why the coherent states give accurate results for the electric field.
\section{A New approach to the understanding of coherent states}
So far we have seen an indepth analysis of the Quantum Harmonic oscillator and its perturbation (with x being shifted by a complex constant $\sqrt{\frac{2\hbar}{\omega}}\alpha$).\\
We have seen that the calculation of electric field operator reproduces the classical results only when we use coherent states in case of the unperturbed oscillator. \\
\textit{Thus we postulate that, there is a transition of the Hamiltonian when we observe the electric field, i.e. the (unperturbed) oscillator(for the electromagnetic field) changes into the perturbed oscillator (with $\hat{q}$ being shifted by a complex constant $\sqrt{\frac{2\hbar}{\omega}}\alpha$).}
\\Mathematically we mean, upon observation:-
$$\hat{H}_{\alpha}{\rightarrow}\hat{H}$$
\\(Where,
\begin{align*}
\hat{H}=\frac{1}{2}(\hat{p}^2+{\omega}^2\hat{q}^2),and\\
\hat{H}_{\alpha}=\frac{\hat{p}^2}{2}+\frac{{\omega}^2}{2}(\hat{q}-\sqrt{\frac{2\hbar}{\omega}}\alpha)^2
\end{align*})
\\Thus in other words we can say that the complex constant $\alpha$  reduces to 0 upon observation.
\\We also make an assumption that this change is almost instantaneous.\\(``When you have eliminated all that is possible, whatever remains must be the truth, no matter how improbable"- this is why we include observation as the cause for the change in the Hamiltonian. Because simply no event can take place without observation. Also other factors often include physical factors that are specific to certain kind of events, thus the equation derived will not be able to sattisfy a generalized case.)
Since, we have assumed that the change in the Hamiltonian is instantaneous, we can use the sudden approximation.
The sudden approximation states that, when the Hamiltonian is changed suddenly, the state cannot catch up with the change and basically remains unchanged.
Thus, in this case we mean that after the transition of $\hat{H}_{\alpha}\rightarrow\hat{H}$ , the ground states will remain at $\ket{\alpha}$.(Since, this transition is instantaneous)
However, The time evolution of the operator $\hat{q}$ will take place in accordance with the Hamiltonian $\hat{H}$ and not $\hat{H}_{\alpha}$, since there is a transition of $\hat{H}_{\alpha}$ to $\hat{H}$.
Thus we have :-
$$\hat{q}(t)=\sqrt{\frac{\hbar}{2\omega}}(\hat{a}e^{-i{\omega}t}+\hat{a}^{\dagger}e^{i{\omega}t})$$
as in equation 5.6.
Now we wil again calculate the expectation value of $\hat{q}(t)$. This time we know, we need to use coherent states $\ket{\alpha}$, (which are the ground states of $\hat{H}_{\alpha}$) for the calculation of expectation value of $\hat{q}(t)$.
Thus, we have:-
\begin{equation}\tag{9.1}
\bra{\alpha}\hat{q}(t)\ket{\alpha}=\sqrt{\frac{\hbar}{2\omega}}({\alpha}e^{-i{\omega}t}+{\alpha}^{*}e^{i{\omega}t})
\end{equation} 
\\(Where ${\alpha}^{*}$ represents the complex conjugate of $\alpha$)
\\Thus the Electric field can be written as :-
$$\hat{E}_{x}(z,t)=2|\alpha|\varepsilon_{0}cos({\omega}t-\theta)\sin{kz}$$
Thus we recover the classical equations as in equation 5.10. However this is not sufficient proof of the fact that there occurs transition of $\hat{H}_{\alpha}\rightarrow\hat{H}$. Thus , let's calculate the expectation value of Electric field in the other energy eigenstates of the Hamiltonian $\hat{H}_{\alpha}$.
Let's represent the n-th eigenstate of the Hamiltonian $\hat{H}_{\alpha}$ by $\ket{\alpha_{n}}$ i.e. $\ket{\alpha_{1}}$ represent the 1st excited state, $\ket{\alpha_{2}}$ represent the 2nd excited state, and so on. However the ground state will be represented by $\ket{\alpha}$ as before.
\subsection{Calculation of the expectation value of Electric field in the 1st excited state of the Hamiltonian \textbf{$\hat{H}_{\alpha}$}}
The state $\ket{\alpha_{1}}$ can be written as:-
\begin{equation}\tag{9.1a}
\ket{\alpha_{1}}=\hat{a}^{\dagger}_{\alpha}\ket{\alpha}
\end{equation}
Thus the expectation value $\hat{q}$ in the states $\ket{\alpha_{1}}$ can be written as:-
\begin{align*}\tag{9.1b}
\bra{\alpha_{1}}\hat{q}\ket{\alpha_{1}}=\bra{\hat{a}^{\dagger}_{\alpha}\alpha}\hat{q}\ket{\hat{a}^{\dagger}_{\alpha}\alpha}=\bra{\alpha}\hat{a}_{\alpha}(\hat{q})\hat{a}^{\dagger}_{\alpha}\ket{\alpha}
\end{align*}
Thus the classical electric field in these states can be written as :-
\begin{align*}\tag{9.1c}
& \bra{\alpha_{1}}\hat{E}_{x}(z,t)\ket{\alpha_{1}}\\\tag{9.1d}
& =\varepsilon_{0}(\bra{\alpha}\hat{a}_{\alpha}\hat{a}\hat{a}^{\dagger}_{\alpha}\ket{\alpha}e^{-i{\omega}t}+\bra{\alpha}\hat{a}_{\alpha}\hat{a}^{\dagger}\hat{a}^{\dagger}_{\alpha}\ket{\alpha}e^{i{\omega}t})\sin{kz}
\end{align*}
Thus to calucalte the expectation value of Electric field, we first should calculate the values of $\bra{\alpha}\hat{a}_{\alpha}\hat{a}\hat{a}^{\dagger}_{\alpha}\ket{\alpha}$ and $\bra{\alpha}\hat{a}_{\alpha}\hat{a}^{\dagger}\hat{a}^{\dagger}_{\alpha}\ket{\alpha}$.
\\\textbf{Calculation of $\bra{\alpha}\hat{a}_{\alpha}\hat{a}^{\dagger}\hat{a}^{\dagger}_{\alpha}\ket{\alpha}$}:-
\begin{align*}
 & \bra{\alpha}\hat{a}_{\alpha}\hat{a}^{\dagger}\hat{a}^{\dagger}_{\alpha} \ket{\alpha}\\
 & =\bra{\alpha}{(\hat{a}-\alpha^{*})\hat{a}^{\dagger}(\hat{a}^{\dagger}-\alpha)}\ket{\alpha}\\
 & =\bra{\alpha}{\hat{a}\hat{a}^{\dagger}\hat{a}^{\dagger}-\alpha^{*}\hat{a}^{\dagger}\hat{a}^{\dagger}-{\alpha}\hat{a}\hat{a}^{\dagger}+|{\alpha}|^{2}\hat{a}^{\dagger}}\ket{\alpha}\\\tag{9.1e}
 & =\bra{\alpha}{\hat{a}\hat{a}^{\dagger}\hat{a}^{\dagger}}\ket{\alpha}-\alpha^{*}\bra{\alpha}{\hat{a}^{\dagger}\hat{a}^{\dagger}}\ket{\alpha}-{\alpha}\bra{\alpha}{\hat{a}\hat{a}^{\dagger}}{\ket{\alpha}}+|{\alpha}|^{2}\bra{\alpha}{\hat{a}^{\dagger}}\ket{\alpha}
\end{align*}
Thus we need to calculate the values of $\bra{\alpha}{\hat{a}\hat{a}^{\dagger}\hat{a}^{\dagger}}\ket{\alpha}$,$\bra{\alpha}{\hat{a}^{\dagger}\hat{a}^{\dagger}}\ket{\alpha}$,$\bra{\alpha}{\hat{a}\hat{a}^{\dagger}}{\ket{\alpha}}$ and $\bra{\alpha}{\hat{a}^{\dagger}}\ket{\alpha}$ to find out the value of  $\bra{\alpha}\hat{a}_{\alpha}\hat{a}\hat{a}^{\dagger}_{\alpha}\ket{\alpha}$.
\\Calculation of $\bra{\alpha}{\hat{a}\hat{a}^{\dagger}\hat{a}^{\dagger}}\ket{\alpha}$:-
\begin{align*}
& \bra{\alpha}{\hat{a}\hat{a}^{\dagger}\hat{a}^{\dagger}}\ket{\alpha}\\
& =\bra{\alpha}{(1+\hat{a}^{\dagger}\hat{a})\hat{a}^{\dagger}}\ket{\alpha}\\
& =\bra{\alpha}{\hat{a}^{\dagger}}\ket{\alpha}+\bra{\alpha}{\hat{a}^{\dagger}\hat{a}\hat{a}^{\dagger}}{\ket{\alpha}}\\
& =\alpha^{*}+\alpha^{*}\bra{\alpha}{\hat{a}\hat{a}^{\dagger}}\ket{\alpha}\\
& =\alpha^{*}+\alpha^{*}(1+|{\alpha}|^2)\\\tag{9.1f}
& =2\alpha^{*}+\alpha^{*}|{\alpha}|^2
\end{align*}\\
Calculation of $\bra{\alpha}{\hat{a}^{\dagger}\hat{a}^{\dagger}}\ket{\alpha}$:-
\begin{align*}\tag{9.1g}
\bra{\alpha}{\hat{a}^{\dagger}\hat{a}^{\dagger}}\ket{\alpha}={\alpha^{*2}}
\end{align*}
Calculation of $\bra{\alpha}{\hat{a}\hat{a}^{\dagger}}{\ket{\alpha}}$ :-
\begin{align}\tag{9.1h}
\bra{\alpha}{\hat{a}\hat{a}^{\dagger}}{\ket{\alpha}}=(1+|\alpha|^2)
\end{align}
Calculation of $\bra{\alpha}{\hat{a}^{\dagger}}\ket{\alpha}$:-
\begin{align*}\tag{9.1i}
\bra{\alpha}{\hat{a}^{\dagger}}\ket{\alpha}=\alpha^{*}
\end{align*}
Putting the values obtained in equations 9.1f,9.1g,9.1h,9.1i in equation 9.1e we have:-
\begin{align*}
& \bra{\alpha}{\hat{a}_{\alpha}\hat{a}\hat{a}^{\dagger}_{\alpha}}\ket{\alpha}\\
& =(2{\alpha}^{*}+\alpha^{*}|\alpha|^2)-\alpha^{*3}-\alpha(1+|\alpha|^2)+\alpha^{*}|\alpha|^2\\\tag{9.1j}
& =2\alpha^{*}(1+|\alpha|^2)-\alpha^{*3}-\alpha(1+|\alpha|^2)
\end{align*}
Similarly we can calculate the value of $\bra{\alpha}{\hat{a}_{\alpha}\hat{a}\hat{a}^{\dagger}_{\alpha}}\ket{\alpha}$. However, there is another shorter way to do that. We use the property that:-
$$(AB)^{\dagger}=B^{\dagger}A^{\dagger}$$.Thus we can write:-
$$(\hat{a}_{\alpha}\hat{a}\hat{a}^{\dagger}_{\alpha})^{\dagger}={\hat{a}_{\alpha}\hat{a}^{\dagger}\hat{a}^{\dagger}_{\alpha}}$$.
So we can state that, $\bra{\alpha}{\hat{a}_{\alpha}\hat{a}^{\dagger}\hat{a}^{\dagger}_{\alpha}}\ket{\alpha}$ is the complex conjugate of $\bra{\alpha}{\hat{a}_{\alpha}\hat{a}\hat{a}^{\dagger}_{\alpha}}\ket{\alpha}$. Thus we have:-
\begin{equation}\tag{9.1k}
\bra{\alpha}{\hat{a}_{\alpha}\hat{a}\hat{a}^{\dagger}_{\alpha}}\ket{\alpha}=2\alpha(1+|\alpha|^2)-\alpha^{3}-\alpha^{*}(1+|\alpha|^2)
\end{equation}
Thus the expectation value of the electric field will be:-
\begin{align*}
& \bra{\alpha_{1}}\hat{E}_x(z,t)\ket{\alpha_{1}}\\
& =\varepsilon_{0}(\bra{\alpha}\hat{a}_{\alpha}\hat{a}\hat{a}^{\dagger}_{\alpha}\ket{\alpha}e^{-i{\omega}t}+\bra{\alpha}\hat{a}_{\alpha}\hat{a}^{\dagger}\hat{a}^{\dagger}_{\alpha}\ket{\alpha}e^{i{\omega}t})\sin{kz}\\
& =\varepsilon_{0}((2\alpha^{*}(1+|\alpha|^2)-\alpha^{*3}-\alpha(1+|\alpha|^2))e^{i{\omega}t}+(2\alpha(1+|\alpha|^2)-\alpha^{3}-\alpha^{*}(1+|\alpha|^2))e^{-i{\omega}t})\sin{kz}\\
& =\varepsilon_{0}((1+|\alpha|^2)(2\alpha^{*}e^{i{\omega}t}+2{\alpha}e^{-i{\omega}t})-(\alpha^{*3}e^{i{\omega}t}+\alpha^{3}e^{i{-\omega}t})-(1+|\alpha|^{2})({\alpha}e^{i{\omega}t}+{\alpha}^{*}e^{-i{\omega}t}))\sin{kz}\\
\intertext{(Putting in $\alpha=|\alpha|e^{i{\theta}}$,we get:-)}\\\tag{9.1l}
& =\varepsilon_{0}(4|\alpha|(1+|\alpha|^2)\cos{({\omega}t-\theta)}-2|\alpha|^3\cos{({\omega}t-3\theta)}+2|\alpha|(1+|\alpha|^2)\cos{({\omega}t+\theta)})\sin{kz}
\end{align*}
Thus we have calculated the expectation value of the electric field operator in the first excited state of the Perturbed oscillator. And the result satisfies our expectation. This time we get a superposition of 3 Electric fields with differing magnitude and having phase diiferences of $2\theta$ with each other.This superposition of Electric fields can again be simplified to get a classical Electric field of the form $A\cos{({\omega}t-\phi)}$ (where $A$ denotes the resultant amplitude of the electric field and $\phi$ represents the resultant phase difference). Thus even in the 1st excited state, the expectation value of the Electric field gives classical results. Thus our assumption holds good for the 1st excited state of the the perturbed Hamiltonian.\\
In a similar way, we can calculate the expectation values of the Electric field under the 2nd, 3rd or n-th excited states of the Perturbed Hamiltonian. However the Calculation of those are very tidious and time consuming. However we can trace a pattern for these solutions.\\
If we calculate the expectation value of the Electric field in the 2nd excited state,we get:-
\begin{align*}\tag{9.1m}
& \bra{\alpha_{2}}\hat{E}_x(z,t)\ket{\alpha_{2}}\\& =(E_{0}\cos{({\omega}t-5\theta)}+ E_{1}\cos{({\omega}t-3{\theta})}++E_{2}\cos{({\omega}t-{\theta})}+E_{3}\cos{({\omega}t+\theta)}+E_{4}\cos{({\omega}t+3\theta)})
\end{align*}
Here $E_{0}$, $E_{1}$, $E_{2}$, $E_{3}$ and $E_{4}$ represent the amplitude of Electric fields. Calculation of expectation value for any other states of the Perturbed oscillator is very,very rigorous. However We can trace out a generalized equation for the expectation value of Electric field in the n-th eigenstate.
As we can see the pattern we can directly generalize our results for the expectation values of the electric field in the n-th eigenstates of the perturbed oscillator.\\
Thus the expectation value of the electric field in the n-th eigenstates can be expressed as:-
\begin{align*}\tag{9.1n}
\bra{\alpha_{n}}{\hat{E}_{x}(z,t)}\ket{\alpha_{n}}=\sum_{r=0}^{2n}E_{r}\cos{({\omega}t-(2n+1-2r){\theta})}
\end{align*}
The $E_{r}$'s in equation 9.1n represent the amplitude of the Electric fields whose superposition make up the total electric field. Let these Electric fields be called as the basic fields. We find that the number of basic fields that make up the total Electric field is (2n+1), where n is the n-th eigenstate of the perturbed oscillator. Thus we have seen that, all the eigenstates of $\hat{H}_{\alpha}$(Hamiltonian for the perturbed oscillator) perfectly reproduces the classical Electric fields, when we consider that the time evolution of the Electric field operator $\hat{E}_{x}(z,t)$ with respect to the $\hat{H}$(the Hamiltonian for the Quantum Harmonic oscillator.).\\ Thus we can now say that our assumption, that the there is an instantaneous transition of $\hat{H}_{\alpha}\rightarrow\hat{H}$ upon observation holds true, since we get classical electric fields for all the energy eigenstates of $\hat{H}_{\alpha}$.
\section{Double Slit experiment- Understanding with the coherent states}
Let's look at the classic double slit experiment and try to explain it with the theory, we have devoloped on the coherent states.
We shall consider the standard set-up for thee double slit experiment, but instead of electrons we shall be sending light through the double slits. We use photomultiplier near the slits to detect the light going through slits, in this set up. \\
We already know the results of the Double-Slit experiment.The results state that when light goes through the slits and a measuring device (like a photomultiplier) is placed near the slits, light acts like a particle, however when we send light through the slits and no measuring device is placed near the slits, we observe an interference pattern on the screen which means it behaves like waves.
\\We will apply the model we have devoloped, so far to explain the results of the double slit experiment.
When no detector is placed, We can easily explain the phenomenan.
When an observation is made on the screen, after light has passed through the silts there is an instantaneous transiltion of the Hamiltonian i.e. $\hat{H}_{\alpha}\rightarrow\hat{H}$. Thus we will have:-
$$\bra{\alpha_{n}}{\hat{E}_{x}(z,t)}\ket{\alpha_{n}}=\sum_{r=0}^{2n}E_{r}\cos{({\omega}t-(2n+1-2r){\theta})}$$,(where $\ket{\alpha_{n}}$ represents the n-th eigenstate of the Hamiltonian $\hat{H}_{\alpha}$)
similarly the magnetic field for the n-th eigenstates, can be represented as:-
$$\bra{\alpha_{n}}{\hat{B}}\ket{\alpha_{n}}=\sum_{r=0}^{2n}B_{r}\sin{({\omega}t-(2n+1-2r){\theta})}$$(the computation for the expectation value of magnetic field is almost similar to that of electric field and can be derived using almost analogous arguments.)
Thus the Electric and magnetic components of light are oscillating, wave like in nature. This contributes to the wave nature of light. Thus we can see the interference patterns on the screen.
\\However whenever a detector is placed near the slits, the same results do not take place. Here, we make an assumption that since observation is done on light in a much more rigorous manner, the transition of Hamiltonian is not instantaneous and it changes, very slowly i.e. $\hat{H}_{\alpha}\rightarrow\hat{H}$(Since we have assumed before that the change in Hamiltonian is due to observation, this assumption is quite intuitive). Thus the sudden approximation does not apply here (since the change is not instantaneous). Here, we can use the adiabatic expression. It states that the during the transition $\hat{H}_{\alpha}\rightarrow\hat{H}$ there is also change of the states from $\ket{\alpha}_{n}$ to $\hat{n}$ i.e. $\ket{\alpha}_{n}\rightarrow\ket{n}$. Thus, now the expectation value of electric field $\hat{E}_{x}(z,t)$ will be as in equation 5.8 i.e.
$$\bra{n}\hat{E_x}(z,t)\ket{n}= \varepsilon_0(\bra{n}\hat{a}\ket{n}e^{-i{\omega}t}+\bra{n}\hat{a}^{\dagger}e^{i{\omega}t}\ket{n})\sin{\textit{kz}}=0$$
Similarly we will alsohave:-
$$\hat{B}=0$$.(After some computation, which is quite similar to the what we have done in the Electric field)
Thus both the Electric and magnetic field of light is now 0. Thus there is no wave like nature of light in this case. Thus we see that light behaves like a particle in this case.
Thus, we have not only properly explained the double slit experiment based on our model, but also in doing so we have prooved that what at first seemed like a vague assumption that observation effects Hamiltonian is actually a pretty strong one.
\section{Conclusions}
 Thus we can conclude that all our assumptions actually hold true i.e.  there is a transition of the Hamiltonian (of the electromagnetic field) when we observe the electric field, i.e. the (unperturbed) oscillator(for the electromagnetic field) changes into the perturbed oscillator (with $\hat{q}$ being shifted by a complex constant $\sqrt{\frac{2\hbar}{\omega}}\alpha$).
This change in the Hamiltonian can also be visualised as a change in the $\hat{q}$ space by a complex constant $\alpha$.
We have derived that the expectation value of the Electromagnetic states in the n-th eigenstate of the Perturbed oscillator gives back the superposition of classical Electric fields i.e.
$$\bra{\alpha_{n}}{\hat{E}_{x}(z,t)}\ket{\alpha_{n}}=\sum_{r=0}^{2n}E_{r}\cos{({\omega}t-(2n+1-2r){\theta})}$$
$$\bra{\alpha_{n}}{\hat{B}}\ket{\alpha_{n}}=\sum_{r=0}^{2n}B_{r}\sin{({\omega}t-(2n+1-2r){\theta})}$$.
Also we have found that in the n-th excited state the Electric field is the suoperposition of 2n+1 basic Electric waves which vary from each other by a phase difference of $2\phi$.
At last we have explained the Double slit experiment using the mathematics we have devoloped so far, rather explaining the phenomenan from a new perspective.\\
Thus now we can safely say, why and how the Electric fields and Magnetic fields in light are described accurately by the coherent states. This is due to the fact, that the classical Electro magnetic theory that was devoloped had the case of observation inbuilt in the equations. However in case of Quantum mechanics, we have to incorporate the phenomenan of observation into the equations and it is not inbuilt. Also Classically the perturbed Hamiltonian bears no significance, since it is non-hermitian. Also we can see why the photon states don't work, Since the electromagnetic field operators are devoloped with respect to classical electromagnetic field theory,we don't incorporate the phenomenan of observation into it.Thus we need to appreciate the freedom and beauty that Quantum mechanics provides to delve further into the realm of electromagnetic field theory.
\section*{Acknowledgements}
The author thanks Mr. Bani kr. Santra and Dr. Krishnendu Mukherjee for extending their help on the subject matter.
\section*{References}
[1] M.Planck, \textit{Verh.Dtsch.Phys.Ges.Berlin}\textbf{2},202(1990);\textit{ibid}. \textbf{2},237(1990) {\\}
[2] W.P. Schleich \textit{Quantum optics in phase space}(Berlin: Wiley-VCH,2001) {\\}
[3] E.C.G. Sudarshan, \textit{Phys. Rev. Lett.} \textbf{10},277(1963) {\\}
[4] R.J. Glauber, \textit{Phys. Rev.} \textbf{131}, 2766(1963) {\\}
[5] R.J. Glauber, \textit{Phys. Rev. Lett} \textbf{10},84(1963) {\\}
[6] W. Heitler, \textit{The Quantum theory of Radiation}(New York: Oxford University Press,1954), pp.76-8 {\\}
[7] S.N.Bose, \textit{Z. Phys.} \textbf{26},178(1924) {\\}
[8] C.L. Mehta, \textit{Phys. Rev. Lett} \textbf{18},752(1967) {\\}
[9] C.L. Mehta and E.C.G.,\textit{Phys. Rev}. \textbf{138},B274(1965)

\end{document}